\documentclass[aip,apl,reprint,noshowpacs,noshowkeys,amssymb,amsmath,amsfonts,bibnotes]{revtex4-2}

\makeatletter
\let\@fnsymbol\@fnsymbol@latex
\@booleanfalse\altaffilletter@sw
\def\frontmatter@makefnmark{\@textsuperscript{\normalfont\@thefnmark}}
\makeatother

\usepackage{graphicx}
\usepackage{theorem}
\usepackage{dcolumn}
\usepackage{longtable}
\usepackage{bm}
\usepackage{accents}
\usepackage{color}
\usepackage[normalem]{ulem}
\usepackage{enumitem}

\newcommand{\ie}{{\textit{i.e.}}}
\newcommand{\eg}{{\textit{e.g.}}}

\newcommand{\TRabi}{T_{\text{Rabi}}}
\newcommand{\Apulse}{A_{\text{pulse}}}
\newcommand{\tpulse}{t_{\text{pulse}}}
\newcommand{\tdelay}{t_{\text{delay}}}
\newcommand{\Pe}{P_{\text{e}}}
\newcommand{\Pp}{P_{\text{p}}}
\newcommand{\Pd}{P_{\text{d}}}
\newcommand{\fp}{f_{\text{p}}}
\newcommand{\fr}{f_{\text{r}}}
\newcommand{\fq}{f_{\text{q}}}
\newcommand{\omegar}{\omega_{\text{r}}}
\newcommand{\omegaq}{\omega_{\text{q}}}
\newcommand{\fqa}{f_{\text{q1}}}
\newcommand{\fqb}{f_{\text{q2}}}
\newcommand{\fd}{f_{\text{d}}}

\newcommand{\fss}{f_{\text{ss}}}
\newcommand{\dfac}{\delta f_{\text{ac}}}

\newcommand{\Vrg}{V_{\text{rg}}}
\newcommand{\Vss}{V_{\text{ss}}}
\newcommand{\Vr}{V_{\text{r}}}

\renewcommand{~}{\,}

\makeatletter
\renewcommand{\figurename}{Fig.}
\renewcommand{\tablename}{Tab.}
\renewcommand{\fnum@figure}[1]{\textbf{\figurename~\thefigure~\textbar} }
\renewcommand{\fnum@table}[1]{\textbf{\tablename~\thetable.} }
\makeatother

\begin{document}

\title{Electron charge qubit with 0.1 millisecond coherence time}

\author{Xianjing Zhou}\thanks{These authors contributed equally to this work.}
\affiliation{Center for Nanoscale Materials, Argonne National Laboratory, Lemont, Illinois 60439, USA\looseness=-1}
\affiliation{Pritzker School of Molecular Engineering, University of Chicago, Chicago, Illinois 60637, USA\looseness=-1}
\author{Xinhao Li}\thanks{These authors contributed equally to this work.}
\author{Qianfan Chen}
\affiliation{Center for Nanoscale Materials, Argonne National Laboratory, Lemont, Illinois 60439, USA\looseness=-1}

\author{Gerwin Koolstra}
\affiliation{Computational Research Division, Lawrence Berkeley National Laboratory, Berkeley, California 94720, USA\looseness=-1}

\author{Ge Yang}
\affiliation{The NSF AI Institute for Artificial Intelligence and Fundamental Interactions, USA\looseness=-1}
\affiliation{Computer Science and Artificial Intelligence Laboratory, Massachusetts Institute of Technology, Cambridge, Massachusetts 02139, USA\looseness=-1}

\author{Brennan Dizdar}
\affiliation{James Franck Institute and Department of Physics, University of Chicago, Chicago, Illinois 60637, USA\looseness=-1}

\author{Yizhong Huang}
\affiliation{Pritzker School of Molecular Engineering, University of Chicago, Chicago, Illinois 60637, USA\looseness=-1}

\author{Christopher S. Wang}
\affiliation{James Franck Institute and Department of Physics, University of Chicago, Chicago, Illinois 60637, USA\looseness=-1}

\author{Xu Han}
\affiliation{Center for Nanoscale Materials, Argonne National Laboratory, Lemont, Illinois 60439, USA\looseness=-1}
\affiliation{Pritzker School of Molecular Engineering, University of Chicago, Chicago, Illinois 60637, USA\looseness=-1}

\author{Xufeng Zhang}
\affiliation{Department of Electrical and Computer Engineering, Northeastern University, Boston, Massachusetts 02115, USA\looseness=-1}

\author{David I. Schuster}\email[Email: ]{dschus@stanford.edu}
\affiliation{Pritzker School of Molecular Engineering, University of Chicago, Chicago, Illinois 60637, USA\looseness=-1}
\affiliation{James Franck Institute and Department of Physics, University of Chicago, Chicago, Illinois 60637, USA\looseness=-1}
\affiliation{Department of Applied Physics, Stanford University, Stanford, California 94305, USA\looseness=-1}

\author{Dafei Jin}\email[Email: ]{dfjin@nd.edu}
\affiliation{Center for Nanoscale Materials, Argonne National Laboratory, Lemont, Illinois 60439, USA\looseness=-1}
\affiliation{Pritzker School of Molecular Engineering, University of Chicago, Chicago, Illinois 60637, USA\looseness=-1}
\affiliation{Department of Physics and Astronomy, University of Notre Dame, Notre Dame, Indiana 46556, USA\looseness=-1}

\date{\today}

\begin{abstract}

Electron charge qubits are compelling candidates for solid-state quantum computing because of their inherent simplicity in qubit design, fabrication, control, and readout. However, all existing electron charge qubits, built upon conventional semiconductors and superconductors, suffer from severe charge noise that limits the coherence time to the order of 1 microsecond. Here, we report our experimental realization of ultralong-coherence electron charge qubits, based upon isolated single electrons trapped on an ultraclean solid neon surface in vacuum. Quantum information is encoded in the motional states of an electron that is strongly coupled with microwave photons in an on-chip superconducting resonator. The measured relaxation time $T_1$ and coherence time $T_2$ are both on the order of 0.1 milliseconds. The single-shot readout fidelity without using a quantum-limited amplifier is 98.1\%. The average single-qubit gate fidelity using Clifford-based randomized benchmarking is 99.97\%. Simultaneous strong coupling of two qubits with the same resonator is demonstrated, as a first step toward two-qubit entangling gates for universal quantum computing. These results manifest that the electron-on-solid-neon (eNe) charge qubits outperform all existing charge qubits to date and rival state-of-the-art superconducting transmon qubits, offering an appealing platform for quantum computing.

\end{abstract}

\maketitle
\pretolerance=9000 

Quantum bits (qubits) are the fundamental building blocks in quantum information processing. A key measure of a qubit's performance is its coherence time, which describes how long a superposition between two quantum states $|0\rangle$ and $|1\rangle$ can persist~\cite{ladd2010}. Among a handful of on-chip solid-state qubits today~\cite{Chatterjee2021, Siddiqi2021}, a coherence time on the order of 0.1~ms or longer has only been achieved in semiconductor quantum-dot and donor qubits based on electron spins~\cite{Stano2022,muhonen2014,veldhorst2014,yoneda2018}, and superconducting transmon and fluxonium qubits based on capacitively and inductively shunted Josephson junctions~\cite{nguyen2019, aaron2021, place2021, wang2022}. By contrast, the coherence time in the traditional semiconductor quantum-dot qubits and superconducting Cooper-pair-box (CPB) qubits based on electron charges (motional states) is at most on the order of $1~\mu$s~\cite{Stano2022,Heinrich2021}. Given that a typical gate time is around 10~ns in such systems, in order to make electron charge qubits serious contenders for quantum computing, it is imperative to increase their coherence time to at least the order of 0.1~ms, that is, a $\gtrsim 10^4$ ratio between the coherence and gate times~\cite{nielsen2010}.

The short coherence time for conventional electron charge qubits is commonly recognized as a result of their high sensitivity to environmental noise, \eg, charge fluctuations in ordinary host materials~\cite{kim2015, yoneda2018}. Nonetheless, if their coherence time can be substantially prolonged, electron charge qubits will possess unique advantages: (i) They can be conveniently designed and fabricated with no need of spin-purified substrates~\cite{veldhorst2014} or patterned micromagnets~\cite{takeda2016fault}, significantly reducing the manufacturing cost~\cite{osman2021}. (ii) They can be fully electrically controlled with no involvement of magnetic fields~\cite{pashkin2009}, intrinsically eliminating the compatibility issues between magnetic fields and superconducting circuits~\cite{samkharadze2016high, kroll2019magnetic}. (iii) They can be individually addressed and readout by microwave photons owing to the much stronger coupling between an electric dipole and electric field than a magnetic dipole and magnetic field~\cite{blais2021, Stano2022}, fundamentally avoiding the complexities of high microwave power~\cite{d2019optimal} or spin-charge conversion~\cite{hu2012strong, takeda2016fault}.

In this paper, we report our experimental realization of unconventional electron charge qubits with 0.1~ms long coherence time, based upon single electrons trapped on a solid neon surface~\cite{zhou2022}. Neon (Ne), as a noble-gas element, is inert against forming chemical bonds with any other elements. In a low-temperature and near-vacuum environment, it spontaneously condenses into an ultrapure semi-quantum solid~\cite{Zavyalov2005} devoid of any two-level-system (TLS) fluctuators, quasiparticles, or dangling bonds that are present in most ordinary materials~\cite{pashkin2009,wilen2021}. Its small atomic polarizability and negligible spinful isotopes make it akin to vacuum with minimal charge and spin noise for electron qubits~\cite{Zavyalov2005,Leiderer2016}. By integrating an electron trap in a superconducting quantum circuit, the charge (motional) states of an electron can be controlled and readout by microwave photons in an on-chip resonator. Our previous demonstration of the electron-on-solid-neon (eNe) qubit platform has shown an appreciable relaxation time $T_1$ of 15~$\mu$s and coherence time $T_2$ of 220~ns~\cite{zhou2022}.

Here we successfully extend both $T_1$ and $T_2$ into 0.1~ms time scale by making three critical advancements: (i) annealing solid Ne to pursue the best surface quality, (ii) stabilizing the electron trapping potential to ensure the lowest ($<10$~Hz) background noise, and (iii) operating the qubit at charge-noise-insensitive (sweet) spots. With 0.1~ms long coherence time, we manage to perform single-shot readout of the qubit states~\cite{mallet2009} and obtain a 98.1~\% readout fidelity without using a quantum-limited amplifier. This is on par with the readout fidelity of the state-of-the-art transmon qubits with a similar amplification chain~\cite{mallet2009,Stefanazzi2022}. We further manage to perform Clifford-based randomized benchmarking~\cite{Knill2008} and obtain an average single-qubit gate fidelity of 99.97~\%, which is well above the fault-tolerance threshold for quantum error correction with surface codes~\cite{Fowler2012}. Moreover, we manage to simultaneously couple two electron qubits with the same resonator, as a first step toward two-qubit entangling gates for universal quantum computing~\cite{divincenzo2000}. These results manifest that the eNe charge qubits outperform all traditional semiconductor and superconducting charge qubits and rival the best superconducting transmon qubits today.

\begin{figure*}[t]
\centerline{\includegraphics[scale=0.95]{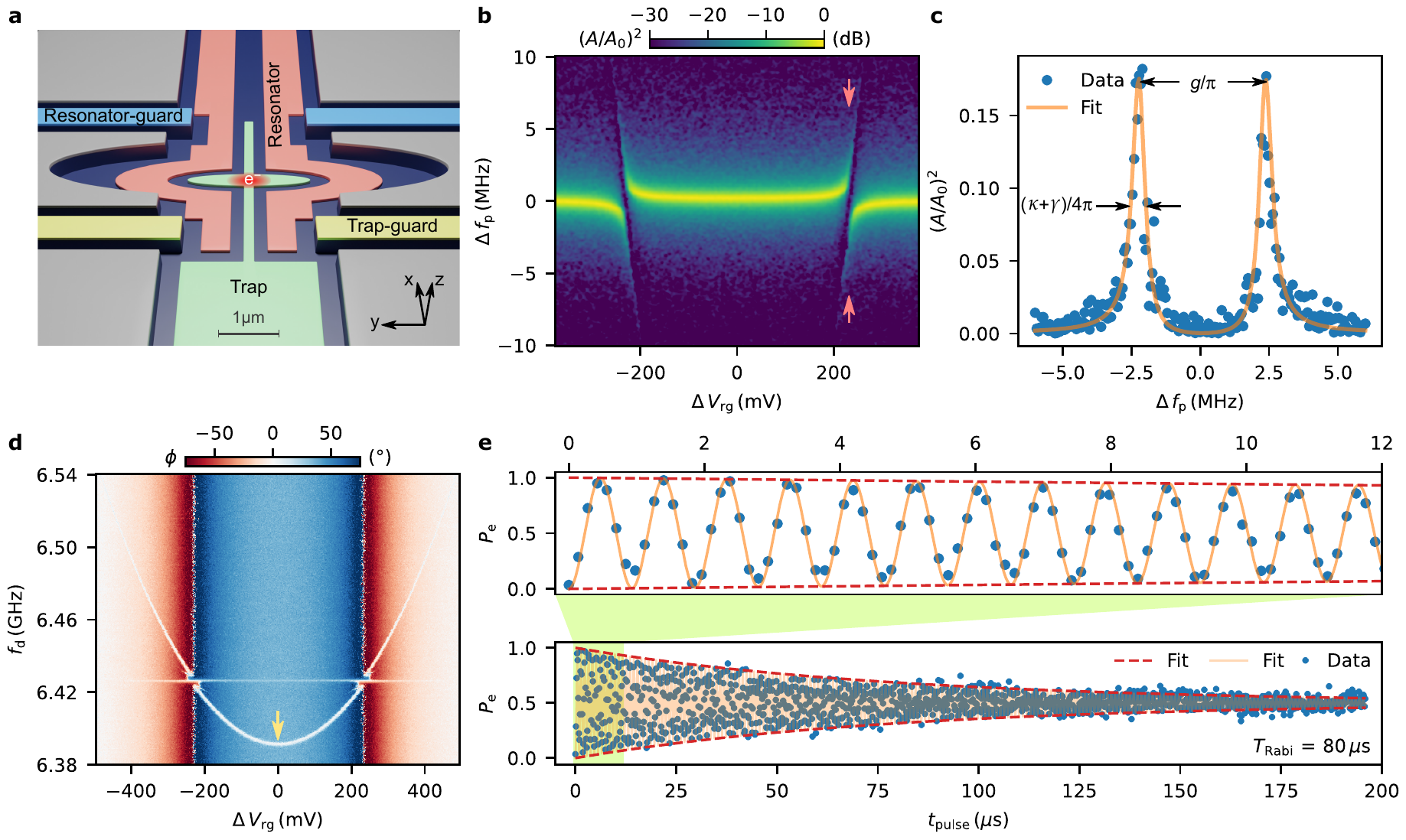}}
\caption{\textbf{Design and properties of an electron-on-solid-neon (eNe) charge qubit}. \textbf{a,} Schematic of the device. A single electron is trapped on a solid Ne surface in the oval region of the channel. Its qubit spectrum is tuned by the dc electrodes around and its motional states in $y$ are coupled with the electric field of microwave photons in the double-stripline resonator. \textbf{b,} Observation of vacuum Rabi splitting. The normalized microwave transmission amplitude $(A/A_{0})^2$ through the resonator is plotted versus the offset probe frequency $\Delta\fp = \fp-\fr$ and the offset resonator-guard voltage $\Delta\Vrg = \Vrg-\Vss$, where $\Vss$ is the value of $\Vrg$ on the sweet spot pointed by the yellow arrow in \textbf{d}. The pink arrows mark the on-resonance condition when $\fq=\fr$. \textbf{c,} Line plot of $(A/A_{0})^2$ versus $\Delta\fp$ at the value of $\Vrg$ indicated by the pink arrows in \textbf{b}, where the qubit and resonator are on resonance, $\fq=\fr$. The two peaks give the coupling strength $g$ and the qubit linewidth $\gamma$ when $\fq=\fr$. \textbf{d,} Two-tone measurement of qubit spectrum. The microwave transmission phase $\phi$ through the resonator is measured at the probe-tone frequency $\fp$ (fixed at the bare resonator frequency $\fp=\fr$) and is plotted against a simultaneously applied drive-tone frequency $\fd$ and $\Delta V_\text{rg}$. The white curve shows the nearly quadratic dependence of qubit frequency $\fq$ on $\Delta V_\text{rg}$. The yellow arrow indicates the minimum called the charge sweet spot. \textbf{e,} Observation of Rabi oscillations in short and long time scales. The excited-state population $\Pe$ is plotted versus the microwave pulse duration $\tpulse$ with a fixed amplitude and qubit frequency. The orange solid curve fits the exponentially decaying sinusoidal oscillations and the red dashed curve fits the exponentially decaying envelop. The fitted Rabi decay time is $T_{\rm{Rabi}}$ = 80~$\mu$s. }  \label{Fig:Structure}
\label{Fig:Spectroscopy}
\end{figure*}

\begin{figure*}[htb]
\centerline{\includegraphics[scale=0.95]{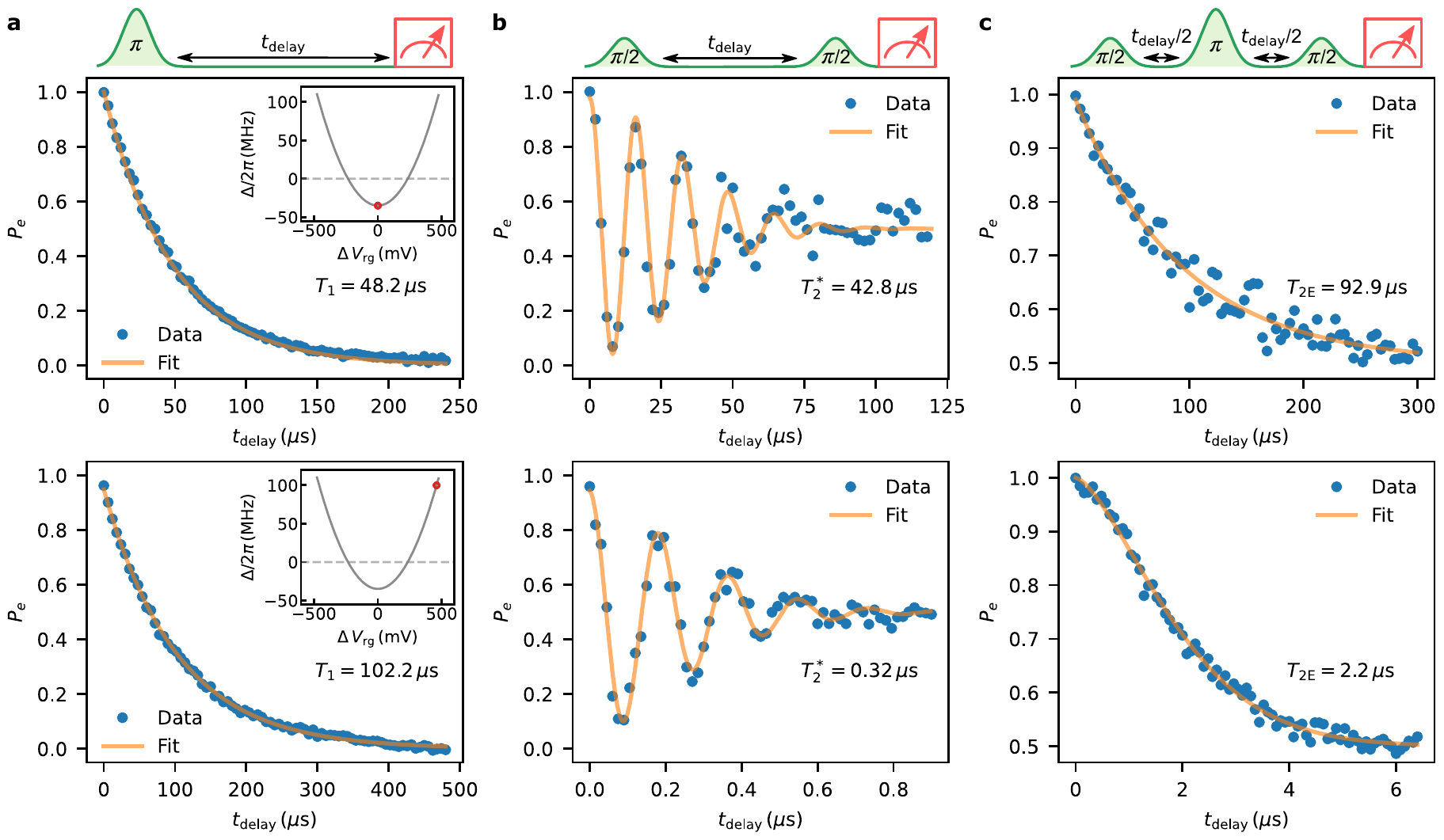}}
\caption{\textbf{Time-domain characterization of an electron-on-solid-neon (eNe) charge qubit}. \textbf{a,} Relaxation time measurements of the qubit on (upper panel) and off (lower panel) the sweet spot. The sweet spot is at the minimum of the qubit spectrum detuned by $\Delta/2\pi = -34.7$~MHz from the bare resonator frequency and the non-sweet spot is chosen at a point with large slope on the qubit spectrum detuned by $\Delta/2\pi = 100$~MHz, as indicated by the circles in the insets. The excited-state population $\Pe$ is plotted versus the delay time $\tdelay$ between the readout pulse and the $\pi$-gate pulse. The fitted relaxation time is $T_1 = 48.2~\mu$s and $102.2~\mu$s, on and off the sweet spot respectively. \textbf{b,} Ramsey-fringe measurements of the qubit on and off the sweet spot. $\Pe$ is plotted versus the delay time $\tdelay$ between two $\pi/2$-gate pulses. The fitted total dephasing time is $T_2^* = 42.8~\mu$s and $0.32~\mu$s, on and off the sweet spot respectively. \textbf{c,} Hahn-echo measurements of the qubit on and off the sweet spot. $\Pe$ is plotted versus the delay time $\tdelay$ between two $\pi/2$-gate pulses separated by a $\pi$-gate pulse in the middle. The fitted total coherence time (with a Hahn echo) is $T_{2\text{E}} = 92.9~\mu$s and $2.2~\mu$s, on and off the sweet spot respectively.}
\label{Fig:Times}
\end{figure*}

\noindent\textbf{Qubit design}

The eNe qubit is situated in an electron trap in a niobium (Nb) superconducting quantum circuit that is fabricated on an intrinsic silicon (Si) substrate, as shown in Fig.~\ref{Fig:Spectroscopy}a. A channel of 3.5~$\mu$m in width and 1~$\mu$m in depth is etched into the substrate. A quarter-wavelength double-stripline microwave resonator runs on the bottom through the channel. A dc electrode, called the trap, also runs on the bottom, but from the other end of the channel into the open end of the resonator. The channel, resonator, and trap are all deformed into oval shapes in the trapping region to accommodate the desired functionalities as described below. On the ground plane outside the channel, four additional dc electrodes, made into two pairs and called the resonator-guards and trap-guards respectively, surround the trapping region. The dc bias voltages applied to these dc electrodes, as well as the resonator with a tuning-fork structure~\cite{koolstra2019}, tune the trapping potential. We ensures the lowest charge noise from our apparatus by using an ultra-stable high-precision digital-to-analog converter (DAC) at room temperature and lowpass filters with 10~Hz cutoff frequency at mK temperature.

The qubit states $|0\rangle$ and $|1\rangle$ are defined by the electron's motional (charge) states, \ie, the ground state $|g\rangle$ and the first excited state $|e\rangle$ respectively, in the $y$-direction across the channel. The electric dipole transition between $|g\rangle$ and $|e\rangle$ strongly couples with the electric field, which points from one stripline to the other, of the microwave photons in the antisymmetric (differential) mode of the resonator~\cite{koolstra2019, zhou2022}. The bare resonator frequency, defined after neon filling but before electron-photon coupling, is $\omegar/2\pi=\fr = 6.4262$~GHz. The resonator linewidth is $\kappa/2\pi = 0.46$~MHz, which is dominated by the input and output photon coupling. All the microwave measurements are done in a transmission configuration through the resonator.

We fill a controlled amount of liquid Ne into the sample cell, using a homemade gas-handling puff system, to wet the channel and quantum circuit at around 26~K. We cool the system down along the liquid-vapor coexistence line and turn the liquid into solid by passing the solid-liquid-gas triple point at the temperature $T_{\text{t}}=24.6$~K and pressure $P_\text{t} = 0.43$~bar~\cite{jacobsen1997thermodynamic}. We hold the temperature at 10~K for 1~--~2 hours to anneal the solid and smooth out the surface~\cite{mugele1992possible}, and then continuously cool down to the base temperature around 10~mK for experiments. The thickness of solid Ne that covers the trapping region is estimated to be tens of nanometers. Electrons are emitted from a heated tungsten filament above the quantum circuit and are trapped on the solid Ne surface under the combined actions of natural surface potential and applied electric potential~\cite{Zavyalov2005, Schuster2010, kawakami2019image, zhou2022}.

\noindent\textbf{Qubit spectroscopy}

We first verify the strong coupling between a trapped single electron and microwave photons in the circuit quantum electrodynamics (cQED) architecture (see Methods). By varying the resonator-guard voltage $\Vrg$ and keeping all other voltages fixed, we tune the qubit frequency $\fq$ across $\fr$. The normalized transmission amplitude $(A/A_0)^2$ through the resonator is plotted in Fig.~\ref{Fig:Spectroscopy}b. Two avoided crossings, known as the vacuum Rabi splitting, can be clearly seen. A line cut in Fig.~\ref{Fig:Spectroscopy}b at the on-resonance condition $\fq=\fr$, marked by the pink arrows, is plotted in Fig.~\ref{Fig:Spectroscopy}c. By fitting the curve with the input-output theory, we obtain the electron-photon (qubit-resonator) coupling strength $g/2\pi = 2.3$~MHz, and the on-resonance qubit linewidth $\gamma/2\pi = 0.36$~MHz. The fact that $g>\kappa>\gamma$ indicates that the qubit and resonator are strongly coupled. In this vacuum Rabi splitting measurement, the average intra-resonator photon number $\bar{n}$ is kept below 1, as can be verified by the ac Stark effect~\cite{schuster2005} (see Methods).

We use two-tone qubit spectroscopy to reveal the qubit spectrum tuned by $V_{\rm{rg}}$, as plotted in Fig.~\ref{Fig:Spectroscopy}d. The dependence of $\fq$ on $\Vrg$ can be identified as the white curve, where the drive frequency $\fd$ hits $\fq$ and induces a sudden phase shift. The spectrum suggests that $\fq$ is nearly a quadratic function of $\Vrg$ and contains a minimum at the so-called charge sweet spot, as indicated by the yellow arrow. On this spot, where $\fq = \fss = 6.3915$~GHz and $\Vrg = \Vss = -270~$mV, the charge qubit is first-order insensitive to the low-frequency charge noise and holds the longest coherence time along the spectrum~\cite{vion2002}.

\begin{figure*}[htb]
\centerline{\includegraphics[scale=0.95]{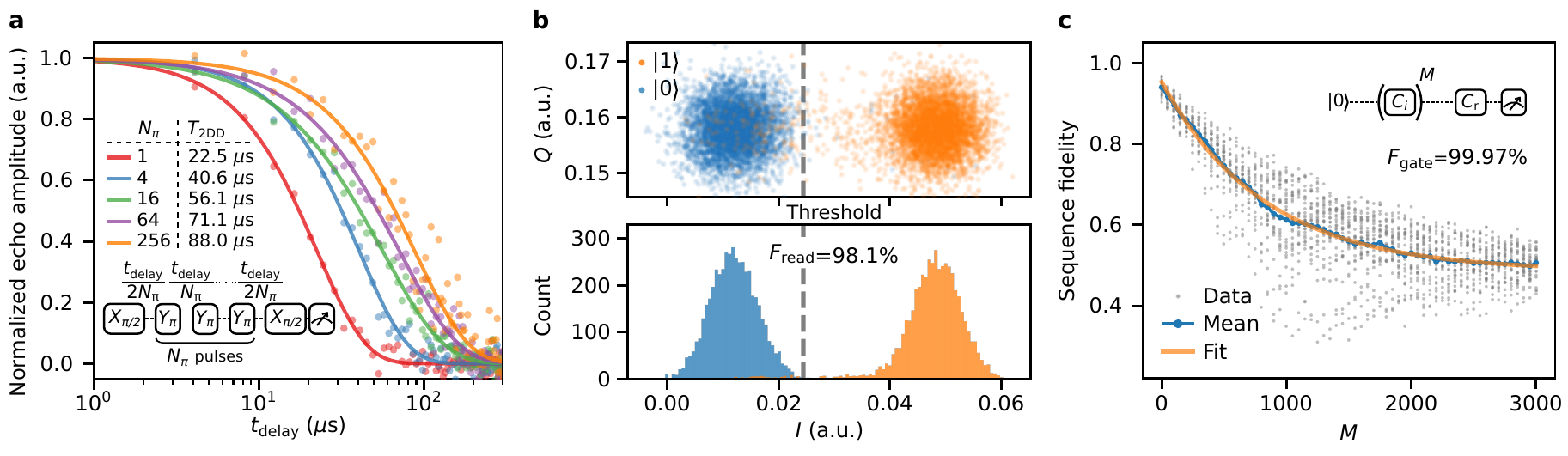}}
\caption{\textbf{Dynamical decoupling, single-shot readout fidelity, and single-qubit gate fidelity of an electron-on-solid-neon (eNe) charge qubit}. \textbf{a,} Normalized echo amplitude versus the total delay time $\tdelay$ for different number $N_\pi$ of CPMG dynamical-decoupling pulses. With $N_\pi=256$, $T_\text{2DD}=88.0~\mu$s is one order of magnitude longer than the $T_2^*=6.1~\mu$s for this qubit. \textbf{b,} Single-shot distribution of 5000 readout values for the qubit states prepared in $|0\rangle$ or $|1\rangle$ using 5~$\mu$s readout pulses and without using a quantum-limited amplifier. The color dots represent the demodulated in-phase ($I$) and quadrature ($Q$) readout signals of the qubit. The largest separation of the two readout clouds is aligned along the $I$ axis through demodulation. The overlapped area yields a single-shot readout fidelity $F_\text{read}=98.1\%$. \textbf{c,} Single-qubit gate fidelity measurement using the Clifford-based randomized benchmarking technique. The Clifford sequence applied to the qubit in the ground state contains $M$ Clifford gates $C_i$ that are randomly chosen from the Clifford group, followed by a recovery Clifford gate $C_\text{r}$, which (ideally) sets the qubit back to the ground state. The mean sequence fidelity at every Clifford depth $M$ are averaged over 30 random sequences (gray dots) with 1000 times of measurement for each sequence. A power-law fit (orange line) of the mean sequence fidelity (blue dotted line) versus $M$ yields an average single-qubit gate fidelity $F_\text{gate}= 99.97\%$.}
\label{Fig:Fidelity}
\end{figure*}

\begin{figure*}[htb]
\centerline{\includegraphics[scale=0.95]{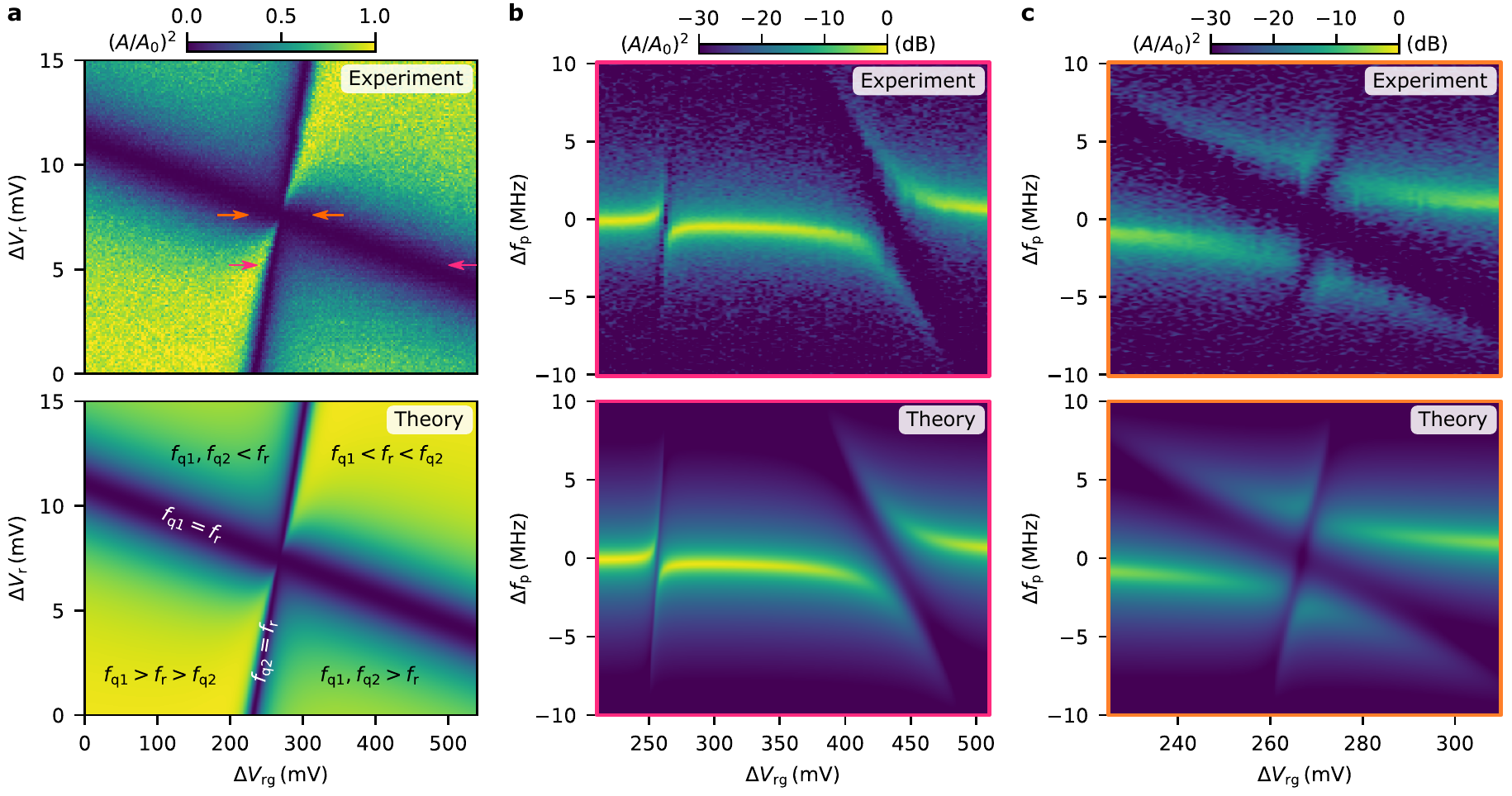}}
\caption{\textbf{Spectroscopic characterization of two electron-on-solid-neon (eNe) charge qubits coupled to a resonator}. Upper row shows the experimental observation and lower row shows the theoretical calculation. The qubit with a larger or smaller coupling strength is labeled as qubit-1 or qubit-2, respectively. \textbf{a,} Spectral tuning of two qubits. The microwave transmission amplitude $(A/A_0)^2$ through the resonator probed at the bare resonator frequency $\fp=\fr$ is plotted against the offset resonator voltage $\Delta \Vr$ and resonator-guard voltage $\Delta \Vrg$. The parameter space ($\Delta \Vrg$, $\Delta \Vr$) is divided into four regions by the two dark lines that correspond to each qubit being individually on resonance with the resonator. The relations between $\fqa$, $\fqb$ and $\fr$ in the four regions and along the two lines are labeled in the lower panel. At the crossing of two dark lines marked by the orange arrows, the two qubits are simultaneously on resonance with the resonator. \textbf{b,} $(A/A_0)^2$ versus $\Delta \fp=\fp - \fr$ and $\Delta \Vrg$ along the line indicated by the pink arrows in \textbf{a}, where the two qubits can be individually on resonance with the resonator. \textbf{c,} $(A/A_0)^2$ versus $\Delta \fp$ and $\Delta \Vrg$ along the line indicated by the orange arrows in \textbf{a}, where the two qubits can be simultaneously on resonance with the resonator.}
\label{Fig:TwoQubits}
\end{figure*}

\noindent\textbf{State control and readout}

We perform real-time state control and readout on the eNe qubit in the dispersive regime. Rabi oscillations~\cite{Krantz2019} are observed by driving the qubit on the sweet spot, using Gaussian-shaped microwave pulses with fixed frequency $\fq$ and amplitude $\Apulse$, and variable pulse duration $\tpulse$. The oscillations of qubit states detected by dispersive readout~\cite{blais2021} (see Methods) are plotted in two (short and long) time scales in Fig.~\ref{Fig:Spectroscopy}e. The Rabi decay time $\TRabi=80~\mu$s is obtained by an exponential fit to the envelope of oscillatory population $\Pe$ in the excited state in the large time scale. Such a long $\TRabi$ indicates both a long relaxation time $T_1$ and a long pure-dephasing time $T_\varphi$, the latter of which is related to the total coherence time $T_2$ via $T_2^{-1} = (2T_1)^{-1} + T_\varphi^{-1}$. Theoretically, in the absence of inhomogeneous broadening and under a strong driving electric field, $\TRabi$ is related to $T_1$ and $T_\varphi$ by $1/T_{\text{Rabi}}=3/(4T_1)+1/(2T_\varphi)$~\cite{Allen1987,bianchetti2009}.

\noindent\textbf{Relaxation and coherence times}

We then find the characteristic times of the eNe qubit, \ie, the relaxation time $T_1$, the total dephasing (Ramsey) time $T_2^*$, and the total coherence time with a Hahn echo $T_{2\text{E}}$. These characteristic times provide key measures of single-qubit performance.

The total relaxation (decay) rate $T_1^{-1} = \Gamma = \Gamma_{\text{R}} + \Gamma_{\text{NR}}$ is the sum of radiative decay rate $\Gamma_{\text{R}}=\kappa g^2/\Delta^2$, which is determined by the Purcell effect~\cite{purcell1946spontaneous, sete2014purcell}, and nonradiative decay rate $\Gamma_{\text{NR}}$. On the sweet spot, the measured $T_1$ is $48.2~\mu$s, as shown in the upper panel of Fig.~\ref{Fig:Times}a. With the known values of $g$, $\kappa$, and $\Delta/2\pi=-34.7$~MHz on the sweet spot, we find a radiative decay time $\Gamma_{\text{R}}^{-1}=78.7~\mu$s and nonradiative decay time $\Gamma_{\text{NR}}^{-1}=125~\mu$s. This suggests that the Purcell-limited radiative decay is the dominant decay channel here. We verify this by purposely moving away from the sweet spot to a point with a larger detuning, $\Delta/2\pi=100$~MHz. It gives an even longer $T_1$ of $102.2~\mu$s, as shown in the lower panel of Fig.~\ref{Fig:Times}a, which agrees with the sum of the estimated $\Gamma_{\text{R}}$ at this detuning and the $\Gamma_{\text{NR}}$ above.

On the sweet spot, the first-order insensitivity of the qubit frequency to the low-frequency charge noise yields exceedingly long total dephasing time $T_2^*$ and total coherence time with a Hahn echo $T_{2\text{E}}$. Our Ramsey-fringe measurement gives a $T_2^* = 42.8~\mu$s, as shown in the upper panel of Fig.~\ref{Fig:Times}b. To our knowledge, this is the longest observed charge-qubit dephasing time, compared with all existing semiconductor quantum-dot and superconducting CPB charge qubits~\cite{wang2022, verjauw2022, Siddiqi2021, Chatterjee2021}. The low-frequency noise can be further suppressed by applying echo pulses. Our Hahn-echo measurement gives a $T_{\rm{2E}} = 92.9~\mu$s that almost equals $2T_1$, as shown in the upper panel of Fig.~\ref{Fig:Times}c. This means that, on the sweet spot, the residual decoherence that cannot be mitigated by Hahn echoes is dominated by relaxation. As a comparison, we purposely move off the sweet spot to a point more sensitive to charge noise at $100$~MHz detuning. The observed $T_2^*$ decreases to $0.32~\mu$s and $T_{\rm{2E}}$ decreases to $2.2$~$\mu$s, as shown in the lower panels of Fig.~\ref{Fig:Times}b and \ref{Fig:Times}c.

For the completeness of our investigation, we slightly change the Ne thickness and make another eNe qubit that has a larger detuning on the sweet spot, $\Delta/2\pi=-288$~MHz, which more strongly suppresses the Purcell-limited radiative decay. This qubit shows a much longer $T_1=82.8~\mu$s but shorter $T_2^*=6.1~\mu$s. However, we successfully utilize the Carr-Purcell-Meiboom-Gill (CPMG) dynamical-decoupling (DD) pulse sequence to push a DD coherence time $T_{2\text{DD}}$ also into 0.1~ms time scale. As shown in Fig.~\ref{Fig:Fidelity}a, with the number of $\pi$ pulses $N_\pi=1$ (equivalent to one echo pulse), $T_{2\text{DD}} = T_{2\text{E}} = 22.5~\mu$s, and with $N_\pi=4$, $16$, $64$, $256$, $T_{2\text{DD}} = 40.6$, $56.1$, $71.1$, $88.0~\mu$s. While systematic analysis of the qubit noise spectra and their correlation with solid Ne quality is underway, the above observations evidence that solid Ne can indeed serve as a superior host material for electron qubits.\\

\noindent\textbf{Readout and gate fidelities}

We then determine the readout and gate fidelities in the eNe qubit system. The qubit with a long $T_1=82.8~\mu$s on the sweet spot allows us to use long ($\gtrsim5\mu$s) readout pulses without relying on a quantum-limited amplifier. Figure~\ref{Fig:Fidelity}b shows the distribution of single-shot readout values for the qubit states prepared in $|0\rangle$ or $|1\rangle$. It yields a single-shot readout fidelity $F_\text{read}=98.1\%$~\cite{Gambetta2007}. This is higher than the reported $94.7\%$ of superconducting transmon qubits with a similar amplification chain~\cite{mallet2009,Stefanazzi2022}.

The single-qubit gate fidelity for the same qubit is found by Clifford-based randomized benchmarking technique~\cite{Knill2008,barends2014superconducting}. In this protocol, a Clifford gate sequence with an increasing number $M$ of random Clifford gates and one recovery gate is applied to the qubit in the ground state. Each gate pulse has a Gaussian shape truncated at 2.5 standard deviation ($\sigma=8$~ns) on each side. A separation time of 20~ns is inserted between every two nearest pulses. The decay of the mean sequence fidelity versus $M$ gives an estimate of the average single-qubit gate fidelity $F_\text{gate}=99.97\%$, as shown in Fig.~\ref{Fig:Fidelity}c, that is well over the threshold for quantum error correction with surface codes~\cite{Fowler2012}.

\noindent\textbf{Two qubits strong coupling}

Beyond the accomplished single-qubit operations, we are able to load two qubits onto the same trap and spectroscopically bring them on and off resonance with the resonator and show strong coupling for each of them. This is the first step to achieve two-qubit entangling gates in a cQED architecture.

We tune the two qubits by concurrently varying two independent parameters: the offset resonator voltage $\Delta \Vr$ and the resonator-guard voltage $\Delta \Vrg$. We call the qubit with larger coupling strength as qubit-1 and the other as qubit-2. Their frequencies $\fqa$ and $\fqb$ have different voltage dependence. The upper row of Fig.~\ref{Fig:TwoQubits} displays the experimental measurements of the normalized transmission amplitude $(A/A_0)^2$ of the resonator at the bare resonance frequency, with variable $\Delta \Vr$ and $\Delta \Vrg$. With obtained qubits properties from the experiments, theoretical calculations based on the Tavis-Cummings model~\cite{tavis1968} and the input-output formalism~\cite{walls2007} (see Methods) show excellent agreement with the experiments, as plotted in the lower row.

The two dark lines in Fig.~\ref{Fig:TwoQubits}a indicate the qubit-resonator on-resonance condition, $\fqa=\fr$ and $\fqb=\fr$, respectively. The parameter space ($\Delta \Vrg$, $\Delta \Vr$) is divided by the two dark lines into four regions: $(\fqa, \fqb) > \fr$, $(\fqa, \fqb) < \fr$, $\fqa>\fr>\fqb$, and $\fqa<\fr<\fqb$. A notable feature is that $(A/A_0)^2$ is smaller in the $(\fqa, \fqb) > \fr$ and $(\fqa, \fqb) < \fr$ two regions, compared with the other regions. In these two regions, both qubits push the resonator frequency in the same direction due to the strong qubits-resonator coupling, resulting a larger resonator frequency shift and thus a smaller transmission amplitude, as verified by the theoretical calculation.

Figure~\ref{Fig:TwoQubits}b shows the system spectrum when the two qubits are individually brought onto resonance with the resonator by a tunable $\Delta \Vrg$ and a fixed $\Delta \Vr=5.2$~mV, as indicated by the magenta arrows in Fig.~\ref{Fig:TwoQubits}a. We can retrieve the coupling strength $g_1/2\pi=3.6$~MHz, $g_2/2\pi=1.8$~MHz, and the qubit linewidth $\gamma_1/2\pi=1.5$~MHz, $\gamma_2/2\pi=1.6$~MHz from the individual vacuum Rabi splitting. Figure~\ref{Fig:TwoQubits}c shows the system spectrum when the two qubits are simultaneously brought onto resonance with the resonator by a tunable $\Delta \Vrg$ and another fixed $\Delta \Vr =7.4$~mV, as indicated by the orange arrows in Fig.~\ref{Fig:TwoQubits}a. At $\Delta \Vrg = 267$~mV, the resonator is simultaneously hybridized with both qubits.

\noindent\textbf{Discussion and outlook}

While our measured coherence time for an eNe qubit has approached 0.1~ms, we believe that it can be further improved by optimizing our device design, drive scheme, and solid-Ne growth procedure. Solely from the material perspective, we do not foresee a practical limit on the charge-qubit coherence time in this system, though theoretical calculation can be done to find out the ultimate decoherence due to thermal phonons or quantum zero-point motion of Ne atoms~\cite{Dykman2003,Chen2022,Pollack1964,Klein1976RareGas}.

The anharmonicity $\alpha$, defined as the frequency difference between the $|g\rangle\rightarrow|e\rangle$ and $|e\rangle\rightarrow|f\rangle$ transitions with $|f\rangle$ being the second excited state, is a critical parameter for the gate time. A larger $\alpha$ ensures a shorter gate time~\cite{Krantz2019}. For our qubit, $\alpha$ is estimated to be greater than $1$~GHz, based on the large detuning range and strong pumping power that we have explored. We were not able to observe a $|g\rangle\rightarrow |f\rangle$ two-photon transition or a $|e\rangle\rightarrow |f\rangle$ one-photon transition after preparing the qubit on $|e\rangle$. We shall note that even for an infinite $\alpha$, which corresponds to an ideal two-level system, the theoretical dispersive shift would be $\chi=g^2/\Delta=-0.152~$~MHz, which is close to our measured $-0.13$~MHz. This is another evidence that our $\alpha$ is very larger, $\alpha\gg|\Delta|$.

While we have managed to simultaneously couple two electron qubits to the same resonator, to realize two-qubit gates in real time in the cQED architecture, we need to push on from the strong resonant regime into the strong dispersive regime. This requires larger $g/\gamma$ and $g/\kappa$~\cite{DavidIsaacSchuster2007}. In light of the observed $\gamma/2\pi\lesssim0.02$~MHz at the charge sweet spot, $g/\gamma$ already satisfies the strong dispersive requirement. To keep fast operations, the resonator linewidth $\kappa$ from the input-output coupling cannot be much smaller than the current $\kappa/2\pi=0.46$~MHz. Therefore, the coupling strength $g$ should be enhanced by about ten times, optimally. This may be fulfilled by using high kinetic-inductance superconducting materials for the on-chip resonator~\cite{shearrow2018atomic,xu2019frequency,han2022superconducting}. Realization of two-qubit gates in the eNe charge qubit platform will establish a further milestone toward universal quantum computing.

\bibliography{LongCoherenceChargeQubitRef}
\newpage
\noindent\textbf{Methods}\\
\noindent\textbf{Qubit-resonator coupled system}

Our electron-photon (qubit-resonator) coupled system adopts a cQED architecture. When the qubit and resonator are uncoupled, the qubit has its bare frequency $\omegaq/2\pi=\fq$. In the presence of a finite coupling strength $g$, the eigenstates of the coupled system are dressed states~\cite{blais2021}. As plotted in Extended Data Fig.1a, in the resonant regime, $\fr=\fq$, the qubit and resonator maximally hybridize, and a vacuum Rabi splitting 2$g$ opens up. In the dispersive regime, the detuning $|\Delta=\omegaq-\omegar|\gg g$, the actual qubit frequency acquires a shift of $(1+2\bar{n})\chi$, in which $\chi$ is called the dispersive shift, $2\bar{n}\chi$ is called the ac Stark shift, and $\bar{n}$ is the average intra-resonator photon number. In this regime, the actual resonator frequency acquires a $+\chi$ or $-\chi$ shift, when the qubit is kept in the excited or ground state, respectively.

\noindent\textbf{ac Stark effect}

We use the two-tone qubit spectroscopy to demonstrate the ac Stark effect and calibrate the average intra-resonator photon number $\bar n$. Keeping $\Vrg = \Vss$ on the sweet spot and the drive power $\Pd$ low, we scan both the drive frequency $\fd$ and the probe power $\Pp$. In this scenario, $\bar{n}$ increases with the increasing $\Pp$ and the qubit frequency $\fq$ shifts under the ac Stark effect~\cite{schuster2005}. Extended Data Fig.1b gives a series of curves of $\phi$ versus $\fd$ with step-increased $\Pp$. The detected $\fq$ is red-shifted by $\dfac \approx -6$~MHz when $\Pp$ (from the vector network analyzer) is increased from $-20$~dBm to 0~dBm. This shift is related to the average intra-resonator photon number $\bar{n}$ by $\dfac = \chi\bar{n}/\pi$~\cite{schuster2005}. Through this measurement, and the measurement of $\chi$ (see below), we know that a probe power $\Pp < -13$~dBm $\approx 0.05$~mW (about $-135~$dBm reaching the sample) corresponds to $\bar{n} < 1$.

\noindent\textbf{Dispersive readout}

The qubit readout follows the standard dispersive readout scheme, where the qubit states are inferred from measuring the phase or amplitude shift of the transmission $S_{21}(\fp)$ through the resonator. The readout is performed with $\bar{n} < 1$. As shown in Extended Data Fig.1c, the resonator frequency is dispersively shifted to $\fr+\chi/2\pi$ or $\fr-\chi/2\pi$, when the qubit is in the excited state $|1\rangle$ or ground state $|0\rangle$. Here on the sweet spot, we have $\chi/2\pi = -0.13$~MHz. The dispersive readout has the highest contrast by fixing the probe frequency $\fp$ at the bare resonance frequency $\fr$ indicated by the gray line, where the phase separation between $|0\rangle$ and $|1\rangle$ is maximal.

\noindent\textbf{Theoretical modeling of two qubits coupling}

The behavior of a system consisting of two qubits interacting with a single-mode resonator can be theoretically described by the Tavis-Cummings (TC) Hamiltonian in the rotating-wave approximation~\cite{tavis1968},
\begin{equation}
	\hat{\mathcal{H}}=\hbar \omega _{\text{r}}\hat{a}^{\dagger }\hat{a}+\sum_{j}\left[ \hbar \omega_{j}\hat{\sigma}^{\dagger }_{j}\hat{\sigma}_{j} +\hbar g_{j}\left( \hat{a}^{\dagger }\hat{\sigma}_{j} + \hat{a}\hat{\sigma}^{\dagger }_{j}\right) \right] , \label{TC model}
\end{equation}
where $j=1,2$ and $\omega _{\text{r}}= 2\pi \fr$ is the resonator frequency, $\hat{a}^{\dagger }$ and $\hat{a}$ are the creation and annihilation operators of microwave photons, $\omega _{j}= 2\pi f_{j}$ is the qubit $j$ frequency, and $g_{j}$ is the coupling strength between the qubit $j$ and the resonator, $\hat{\sigma}_{j} =\hat{\sigma}_{j}^{x}-i\hat{\sigma}_{j}^{y}$ and $\hat{\sigma}^{\dagger }_{j}=\hat{\sigma}_{j}^{x}+i\hat{\sigma}_{j}^{y}$ are the ladder operators acting on the qubit $ j $.

Using the quantum master equation and input-output theory~\cite{walls2007}, assisted with the known quantities directly measured from experiments, such as the resonator linewidth $\kappa$, we can theoretically calculate all the normalized transmission amplitude $(A/A_0)^2$ through the resonator versus the qubit frequency $f_{j}$ and probe frequency $\fp$. The only phenomenological modeling necessary is the relations between the qubit frequencies and applied voltages, which we assume to be linear functions.\\

\noindent\textbf{Data availability}\\
The raw data that support the findings of this study are available from the corresponding author upon reasonable request.\\

\noindent\textbf{Code availability}\\
The codes used to perform the experiments and to analyze the data in this work are available from the corresponding author upon reasonable request.

\acknowledgements

Work performed at the Center for Nanoscale Materials, a U.S. Department of Energy Office of Science User Facility, was supported by the U.S. DOE, Office of Basic Energy Sciences, under Contract No. DE-AC02-06CH11357. D.J., X.H., X.L., and Q.C. acknowledge support from Argonne National Laboratory Directed Research and Development (LDRD). D.J. and X.Zhou acknowledge support from the Julian Schwinger Foundation for Physics Research. This work was partially supported by the University of Chicago Materials Research Science and Engineering Center, which is funded by the National Science Foundation under award number DMR-2011854. This work made use of the Pritzker Nanofabrication Facility of the Institute for Molecular Engineering at the University of Chicago, which receives support from SHyNE, a node of the National Science Foundation National Nanotechnology Coordinated Infrastructure (NSF NNCI-1542205). D.I.S. and B.D. acknowledge support from the National Science Foundation DMR grant DMR-1906003. D.I.S. and C.S.W. acknowledge support from the U.S. Department of Energy, Office of Science, National Quantum Information Science Research Centers. G.Y. acknowledges supports from the National Science Foundation under Cooperative Agreement PHY-2019786 (the NSF AI Institute for Artificial Intelligence and Fundamental Interactions). D.J. thanks Anthony J. Leggett for inspiring discussions. The qubit manipulation and measurement in this work utilized the highly efficient and effective OPX+, Octave, and QDAC-II made by Quantum Machines and QDevil.

\newpage

\makeatletter
\renewcommand{\figurename}{Extended Data Fig.}
\renewcommand{\tablename}{Tab.}
\renewcommand{\fnum@figure}[1]{\textbf{\figurename~\thefigure~\textbar} }
\renewcommand{\fnum@table}[1]{\textbf{\tablename~\thetable.} }
\makeatother

\setcounter{figure}{0}

\begin{figure*}[htb]
	\centerline{\includegraphics[scale=0.95]{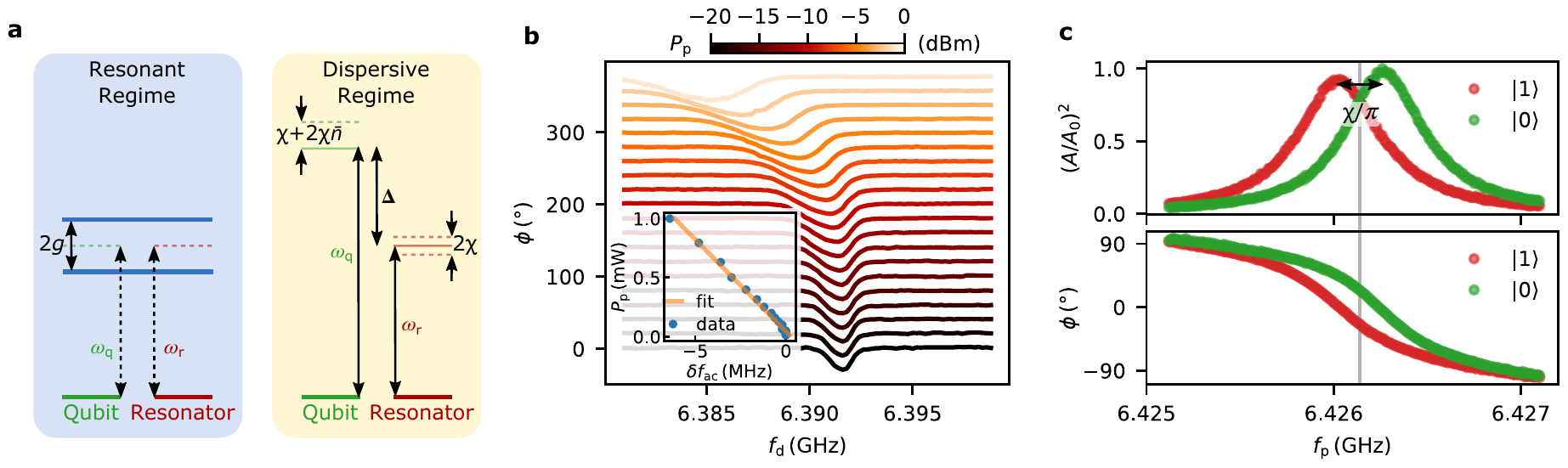}}
	\caption{\textbf{Qubit-resonator coupled spectrum, ac Stark shift, and dispersive shift}. \textbf{a}, Schematic of the qubit-resonator coupled spectrum. $\omegaq=2\pi\fq$ is the bare qubit frequency, $\omegar=2\pi\fr$ is the bare resonator frequency, and $g$ is the coupling strength. In the resonant regime, $\fr=\fq$, the qubit and resonator hybridize and a vacuum Rabi splitting 2$g$ opens up. In the dispersive regime, the detuning $|\Delta=\omegaq-\omegar|\gg g$, the actual qubit frequency exhibits the dispersive shift $\chi$ and the ac Stark shift $2\chi\bar{n}$, in which $\bar{n}$ is the average intra-resonator photon number, whereas the actual resonator frequency exhibits a $+\chi$ or $-\chi$ shift, when the qubit is kept in the excited or ground state, respectively. \textbf{b}, Observation of the ac Stark shift. The transmission phase $\phi$ at $\fp=\fr$ is plotted versus $\fd$ and probe power $\Pp$, when the qubit is on the sweet spot in Fig.~\ref{Fig:Spectroscopy}d. With increasing $\Pp$, the qubit frequency is red-shifted because of the ac Stark effect. In the inset, the frequency shift $\dfac$ shows a linear dependence on $\Pp$ (equivalent to the average intra-resonator photon number $\bar n$). \textbf{c}, Measurement of the state-dependent dispersive shift. Normalized transmission amplitude $(A/A_0)^2$ (top) and phase $\phi$ (bottom) are plotted versus the probe frequency $\fp$ when the qubit is in the ground state $|0\rangle$ or excited state $|1\rangle$. The grey line corresponds to $\fp=\fr$, where $\fr$ is the bare resonator frequency. The measured dispersive shift is $\chi/2\pi=-0.13$~MHz.}
	\label{Fig:Extended}
\end{figure*}

\end{document}